# High-Precision Temperature Estimation Based on Magnetic Nanoparticles Dominated by Brownian Relaxation under Combined AC and DC Magnetic Fields

Zhongzhou Du, Wenze Zhang, Yi Sun, Na Ye, Yong Gan, Pengchao Wang, Xinwei Zhang, Yuanhao Zhang, Shijie Han, Haochen Zhang, Haozhe Wang, Wenzhong Liu, *Member, IEEE*, and Takashi Yoshida, *Member, IEEE*

*Abstract*—**Brownian relaxation is one of the primary mechanisms that allows magnetic nanoparticles (MNPs) to convert magnetic energy into thermal energy under an excitation magnetic field. Accurately characterizing the MNPs' magnetization dynamics dominated by Brownian relaxation is crucial for achieving high-precision temperature estimation. However, the lack of a readily applicable analytical expression remains a major obstacle to the advancement of magnetic nanoparticle hyperthermia (MNPH). In this paper, the perturbation method was applied to derive analytical expressions from the Fokker-Planck equation, which characterized MNPs' magnetization behaviors under the AC and DC magnetic fields. Numerical simulations were conducted to validate the accuracy of the analytical expressions and to explore the correlation between temperature and the magnetization response. Then, a temperature analysis model based on magnetization harmonics was constructed. The first and second harmonic ratios and first harmonic phase were used to calculate MNPs' temperature, respectively. The experimental results demonstrate that within 310 K to 320 K, the estimation error of the temperature using the amplitude ratio of the first to second harmonics is below 0.0151 K, while the error using the first harmonic phase is below 0.0218 K. The derived analytical expressions are expected to enhance the accuracy of MNP-based temperature measurements and facilitate their broader applications in MNPH and MNP imaging.**

## I. INTRODUCTION

MAGNETIC nanoparticle hyperthermia (MNPH) is emerging as a promising technique in cancer therapy. It provides a minimally invasive approach for localized heating of tumor tissues [1][2][3]. In this method, magnetic nanoparticles (MNPs) are injected into cancerous tissues and subsequently subjected to an excitation magnetic field [4], where they generate heat by transforming magnetic energy into heat through mechanisms such as Brownian rotation and magnetic hysteresis loss. MNPs can induce cancer cell death by locally raising the temperature to around 319 K, thereby minimizing collateral damage to non-diseased cells and tissues. However, the therapeutic efficacy and safety of MNPH are significantly influenced by precise thermal regulation within the treatment zone. Precise in vivo temperature measurement can facilitate the advancement of hyperthermia technology [9][10][11].

In previous studies [12][13][14], temperature measurements based on harmonics were typically solved using the Taylor expansion of the Langevin function. This method is applicable only to low frequencies (quasi-static), where the Brownian relaxation is insignificant. It fails to capture the dynamic behavior of MNPs under high excitation frequencies [15]. The MNPs' magnetization dynamics are significantly influenced by Brownian relaxation, a phenomenon that is invariably present under an excitation magnetic field. The Fokker-Planck equation provides a comprehensive theoretical framework for characterizing the Brownian relaxation-dominated MNP magnetization behaviors [16]. However, due to the mathematical complexity of its solution, previous studies have been unable to derive exact analytical expressions for the Fokker-Planck equation dominated by Brownian relaxation in this context. As a result, many studies have relied on empirical fitting methods to approximate the magnetization and magnetic harmonics of MNPs [15][17][18][19]. These studies were the first to explore the influence of relaxation mechanism on MNPs magnetization behaviors, which has promoted the advancement of MNPH. But the fitting-based approaches lack

This work was supported by National Key Research and Development Program of China under Grant 2022YFE0107500, Natural Science Foundation of Henan under Grant 252300420388, the Science and Technology Program of Henan under Grant 252102120209, and China-Belt and Road Joint Laboratory on Measurement and Control Technology under Grant MCT202304. *(Corresponding author: Yi Sun and Wenzhong Liu.)*

Zhongzhou Du, Wenze Zhang, Na Ye, Yong Gan, Pengchao Wang, Xinwei Zhang, and Yuanhao Zhang are with the School of Computer S cience and Technology, Zhengzhou University of Light Industry, Zhengz hou 450001, China (e-mail: duzhongzhou1983@gmail.com; qizhang9608 @gmail.com; yena@zzuli.edu.cn; yonggan985@gmail.com; wangpengchao 0507@gmail.com; xinweizhang0406@gmail.com; zyh5875@gmail.com).

Yi Sun, Haochen Zhang, Haozhe Wang, and Takashi Yoshida are with the Department of Electrical and Electronic Engineering, Kyushu University, Fukuoka 819-0395, Japan (e-mail: sunyi3064@gmail.com; zhang.haochen.048@s.kyushu-u.ac.jp; wang.haozhe.726@s.kyushu-u.ac.jp; t_yoshi@ees.kyushu-u.ac.jp).

Shijie Han and Wenzhong Liu are with the School of Artificial Intelligence and Automation, Huazhong University of Science and Technology, Wuhan 430074, China (e-mail: hanshijie@hust.edu.cn; lwz7410@hust.edu.cn).

generalizability, and often require extensive calibration. This limits their reliability in accurately describing underlying physical mechanisms, including Brownian relaxation, magnetic hysteresis, and Néel relaxation [20]. The absence of analytical expressions limits the interpretability and predictive capability of these models, particularly when applied to biologically relevant environments with varying viscosity [21][22], magnetic field strength, and nanoparticle properties. Consequently, this significantly hinders the accuracy and practical applicability of temperature estimation techniques in MNPH applications [1][2].

In this paper, a more comprehensive theoretical framework for characterizing the Brownian relaxation-dominated magnetization behaviors in MNPs is proposed. Specifically, we derived analytical expressions for magnetization and magnetic harmonics from the Fokker-Planck equation dominated by Brownian relaxation using a perturbation method. This method could directly calculate the accurate analytical expressions of magnetization and magnetic harmonics under the combination of AC and DC excitation magnetic fields without relying on empirical fitting methods. The feasibility of the analytical expressions for magnetization responses and magnetic harmonics was validated through numerical simulations. Furthermore, a temperature analysis model was developed based on the harmonic expressions. For the experimental portion of this study, MNPs' temperature was calculated using the harmonic amplitude ratio and the phase, respectively. The model proposed in this paper may greatly promote the advancement of MNPH and MNP imaging.

## II. MODEL AND METHOD

For Brownian relaxation-dominated MNPs, their magnetization dynamics can be precisely characterized by the Fokker-Planck equation [20][23],

$$2\tau_{B,0}\frac{\partial W(\theta,t)}{\partial t} = \frac{1}{\sin\theta}\frac{\partial}{\partial\theta}\left\{\sin\theta\left[\zeta W(\theta,t)\sin\theta + \frac{\partial W(\theta,t)}{\partial\theta}\right]\right\} \quad (1)$$

In Eq. (1), $\tau_{B,0} = \pi\eta d_h{}^3/2k_BT$ is Brownian relaxation time, $\eta$ represents the viscosity coefficient, and $d_h$ represents the hydrodynamic size, $k_B$ represents The Boltzmann constant, and $T$ represents the absolute temperature. $\xi = mH/k_BT$, where magnetic moment $m = M_sV$, $M_s$ represents saturation magnetization, $V = \pi d_c{}^3/6$ represents the particle volume. $H = \mu_0H_0\cos(\omega t) + \mu_0H_1$ represents the combine excitation field of AC and DC, $\mu_0$ represents the magnetic permeability of vacuum, $H_0$ and $H_1$ respectively represents the amplitude of AC field and DC field. $\omega = 2\pi f$ represents the angular frequency, $f$ represents the excitation frequency. $\theta$ is the angle between the applied field ($H$) and magnetic moment ($m$), and $W(\theta,t)$ represents the distribution function of $\theta$.

The Fokker-Planck equation can be solved using the Legendre polynomial expansion, resulting in a numerical iterative algorithm for the Brownian relaxation process. The Brownian relaxation coefficients can be determined using the following formula:

$$\frac{2\tau_{B,0}}{n(n+1)}\dot{a_n} + a_n = \zeta\left(\frac{a_{n-1}}{2n-1} - \frac{a_{n+1}}{2n+3}\right) \quad (2)$$

Our analysis is restricted to the condition of a weak applied field ($\xi << 1$). Thus, the nonlinear responses are accessible through perturbation methods, which are justified under this approximation. We set:

$$a_n(t) = a_n^0(t) + a_n^1(t) + a_n^2(t) + a_n^3(t) + \cdots \quad (3)$$

Next, combining Eqs. (2) and (3), Eq. (2) can be rewritten as the following perturbation equation for $n > 0$:

$$\frac{2\tau_{B,0}}{n(n+1)}\dot{a_n^i}(t) + a_n^i(t) = \zeta(t)\left(\frac{a_{n-1}^{i-1}}{2n-1} - \frac{a_{n+1}^{i-1}}{2n+3}\right), \ i = 1,2,3,4,5,\cdots \quad (4)$$

where $a_0^0(t) = 1/3$, and $a_n^0(t) = 0$.

Here, we present the solution of Eq. (4), which enables the calculation of $a_1(t)$, when $i = 1$, Eq. (4) can be written as:

$$\frac{2\tau_{B,0}}{n(n+1)}\dot{a_n^1}(t) + a_n^1(t) = 0, \ n > 1 \quad (5)$$

For $n > 1$, $a_n^1(t)$ is equal to 0. When $n = 1$, and $i = 1$, Eq. (4) is re-expressed as the following differential equation:

$$\tau_{B,0}\dot{a_1^1}(t) + a_1^1(t) = \frac{\zeta(t)}{3} \quad (6)$$

Therefore, $a_1^1(t)$ can be obtained by solving Eq. (6).

$$a_1^1(t) = \frac{mH_0\omega\tau_{B,0}}{3k_BT(1+\omega^2\tau_{B,0}^2)}\cos(\omega t)$$
$$+ \frac{mH_0}{3k_BT(1+\omega^2\tau_{B,0}^2)}\sin(\omega t) \quad (7)$$
$$+ \frac{mH_1}{3k_BT}$$

The function $a_2^2(t)$ can be obtained by solving Eq. (4) for the case where $i = 2$, which gives:

$$\frac{2\tau_{B,0}}{n(n+1)}\dot{a_n^2}(t) + a_n^2(t) = \zeta(t)\left(\frac{a_{n-1}^1(t)}{2n-1} - \frac{a_{n+1}^1(t)}{2n+3}\right) \quad (8)$$

Since $a_n^2(t) = 0$ (for $n \neq 2$), the only contribution to Eq. (8) comes from the term with $n = 2$. Setting $n = 2$ and $i = 2$, Eq. (8) becomes:

$$\frac{\tau_{B,0}}{3}\dot{a_2^2}(t) + a_2^2(t) = \zeta(t)\left(\frac{a_1^1(t)}{3} - \frac{a_3^1(t)}{7}\right) \quad (9)$$

Eq. (9) may be integrated by quadrature with $a_2^2(-\infty) = 0$ to yield,

$$a_2^2(t) = \frac{mH_0}{k_BT\tau_{B,0}}\int_{-\infty}^t e^{3(t'-t)/\tau_{B,0}}\cos(\omega t')a_1^1(t')dt'$$
$$+ \frac{mH_1}{k_BT\tau_{B,0}}\int_{-\infty}^t e^{3(t'-t)/\tau_{B,0}}a_1^1(t')dt' \quad (10)$$

Therefore, $a_2^2(t)$ can be obtained;

$$a_2^2(t) = +\frac{2m^2 H_0 \omega \tau_{B,0}(\omega^2 \tau_{B,0}^2 + 3)}{3k_B^2 T^2 (1 + \omega^2 \tau_{B,0}^2)(9 + \omega^2 \tau_{B,0}^2)}\cos(\omega t)$$

$$+\frac{m^2 H_0 \omega \tau_{B,0}(\omega^2 \tau_{B,0}^2 + 5)}{3k_B^2 T^2 (1 + \omega^2 \tau_{B,0}^2)(9 + \omega^2 \tau_{B,0}^2)}\sin(\omega t)$$

$$-\frac{m^2 H_0^2 (2\omega^2 \tau_{B,0}^2 - 3)}{6k_B^2 T^2 (1 + \omega^2 \tau_{B,0}^2)(9 + 4\omega^2 \tau_{B,0}^2)}\cos(2\omega t) \quad (11)$$

$$+\frac{5m^2 H_0^2 \omega \tau_{B,0}}{6k_B^2 T^2 (1 + \omega^2 \tau_{B,0}^2)(9 + 4\omega^2 \tau_{B,0}^2)}\sin(2\omega t)$$

$$+\frac{m^2 (2H_1^2 \omega^2 \tau_{B,0}^2 + H_0^2 + 2H_1^2)}{18k_B^2 T^2 (1 + \omega^2 \tau_{B,0}^2)}$$

Given that $a_n^1(t) = 0$ $(n > 1)$ and $a_n^2(t) = 0$ $(n \neq 2)$, we can obtain the result for $i = 3$ from Eq. (4) as follows:

$$\frac{2\tau_{B,0}}{n(n+1)}\dot{a}_n^3(t) + a_n^3(t) = \zeta(t)\left(\frac{a_{n-1}^2(t)}{2n-1} - \frac{a_{n+1}^2(t)}{2n+3}\right) \quad (12)$$

This can again be integrated by quadrature with $a_1^3(-\infty) = 0$. The result of the calculation is given as:

$$\tau_{B,0}\dot{a}_1^3(t) + a_1^3(t) = \zeta(t)\left(-\frac{a_2^2(t)}{5}\right) \quad (13)$$

The calculated result for $a_1^3(t)$ is as follows:

$$a_1^3(t) = \alpha^{'1}_{13}\cos(\omega t) + \alpha^{''1}_{13}\sin(\omega t)$$
$$+\alpha^{'2}_{13}\cos(2\omega t) + \alpha^{''2}_{13}\sin(2\omega t) \quad (14)$$
$$+\alpha^{'3}_{13}\cos(3\omega t) + \alpha^{''3}_{13}\sin(3\omega t)$$
$$+\alpha^R_{13}$$

where:

$$\alpha^{'1}_{13} = -(H_0 m^3 \omega \tau_{B,0}(8H_1^2 \omega^6 \tau_{B,0}^6 + H_0^2 \omega^4 \tau_{B,0}^4 + 170H_1^2 \omega^4 \tau_{B,0}^4$$
$$+30H_0^2 \omega^2 \tau_{B,0}^2 + 678H_1^2 \omega^2 \tau_{B,0}^2 + 189H_0^2 + 756H_1^2))$$
$$/(90k_B^3 T^3 (9 + \omega^2 \tau_{B,0}^2)(9 + 4\omega^2 \tau_{B,0}^2)(1 + \omega^2 \tau_{B,0}^2)^2) \quad (15)$$

$$\alpha^{''1}_{13} = (H_0 m^3 (32H_1^2 \omega^6 \tau_{B,0}^6 + 13H_1^2 \omega^4 \tau_{B,0}^4 + 56H_1^2 \omega^4 \tau_{B,0}^4$$
$$+90H_0^2 \omega^2 \tau_{B,0}^2 - 468H_1^2 \omega^2 \tau_{B,0}^2 - 243H_0^2 - 972H_1^2))$$
$$/(180k_B^3 T^3 (9 + \omega^2 \tau_{B,0}^2)(9 + 4\omega^2 \tau_{B,0}^2)(1 + \omega^2 \tau_{B,0}^2)^2) \quad (16)$$

$$\alpha^{'2}_{13} = (H_0^2 H_1 m^3 (8\omega^6 \tau_{B,0}^6 + 62\omega^4 \tau_{B,0}^4 + 153\omega^2 \tau_{B,0}^2 - 81))$$
$$/(30k_B^3 T^3 (9 + \omega^2 \tau_{B,0}^2)(9 + 4\omega^2 \tau_{B,0}^2)(1 + \omega^2 \tau_{B,0}^2)(1 + 4\omega^2 \tau_{B,0}^2)) \quad (17)$$

$$\alpha^{''2}_{13} = (-2H_0^2 H_1 m^3 \omega \tau_{B,0}(4\omega^4 \tau_{B,0}^4 + 22\omega^2 \tau_{B,0}^2 + 63))/(15k_B^3 T^3$$
$$(9 + \omega^2 \tau_{B,0}^2)(9 + 4\omega^2 \tau_{B,0}^2)(1 + \omega^2 \tau_{B,0}^2)(1 + 4\omega^2 \tau_{B,0}^2)) \quad (18)$$

$$\alpha^{'3}_{13} = (H_0^3 m^3 (17\omega^2 \tau_{B,0}^2 - 3))/(60k_B^3 T^3$$
$$(1 + 9\omega^2 \tau_{B,0}^2)(9 + 4\omega^2 \tau_{B,0}^2)(1 + \omega^2 \tau_{B,0}^2)) \quad (19)$$

$$\alpha^{''3}_{13} = (H_0^3 m^3 \omega \tau_{B,0}(3\omega^2 \tau_{B,0}^2 - 7))/(30k_B^3 T^3$$
$$(1 + 9\omega^2 \tau_{B,0}^2)(9 + 4\omega^2 \tau_{B,0}^2)(1 + \omega^2 \tau_{B,0}^2)) \quad (20)$$

$$\alpha^R_{13} = (-H_1 m^3 (2H_1^2 \omega^4 \tau_{B,0}^4 + 7H_0^2 \omega^2 \tau_{B,0}^2 + 20H_1^2 \omega^2 \tau_{B,0}^2 + 27H_0^2$$
$$+18H_1^2))/(90k_B^3 T^3 (1 + \omega^2 \tau_{B,0}^2)^2 (9 + \omega^2 \tau_{B,0}^2)) \quad (21)$$

In a similar manner, we can obtain $a_1^i$ ($i = 5, 7, 9, 11, \ldots$) using the above procedure. Thus, the analytical expressions for the magnetization $M$ dominated by Brownian relaxation can be expressed as follows:

$$M = \sum_{j=1}^{\infty} A_j \sin(j\omega t + \varphi_j) + \sum_{j=1}^{\infty} \alpha^R_{1(2j-1)} \quad (22)$$

Based on the analytical expressions of the magnetization response, the harmonic expressions can be derived as:

$$\begin{cases} C_1 = A_1 \sin(\omega t + \varphi_1) \\ C_2 = A_2 \sin(2\omega t + \varphi_2) \\ C_3 = A_3 \sin(3\omega t + \varphi_3) \end{cases} \quad (23)$$

where $C_1$, $C_2$ and $C_3$ represent the 1st, 2nd, and 3rd harmonics respectively. $A_1$, $A_2$, and $A_3$ represent the 1st, 2nd, and 3rd harmonic amplitudes. $\varphi_1$, $\varphi_2$, and $\varphi_3$ represent the 1st, 2nd, and 3rd harmonic phases. where

$$\begin{cases} A_1 = \sqrt{(\alpha^{'1}_{11} + \alpha^{'1}_{13} + \cdots)^2 + (\alpha^{''1}_{11} + \alpha^{''1}_{13} + \cdots)^2} \\ A_2 = \sqrt{(\alpha^{'2}_{13} + \cdots)^2 + (\alpha^{''2}_{13} + \cdots)^2} \\ A_3 = \sqrt{(\alpha^{'3}_{13} + \cdots)^2 + (\alpha^{''3}_{13} + \cdots)^2} \end{cases} \quad (24)$$

$$\begin{cases} \varphi_1 = -\arctan\left((\alpha^{'1}_{11} + \alpha^{'1}_{13} + \cdots)/(\alpha^{''1}_{11} + \alpha^{''1}_{13} + \cdots)\right) \\ \varphi_2 = -\arctan\left((\alpha^{'2}_{13} + \cdots)/(\alpha^{''2}_{13} + \cdots)\right) \\ \varphi_3 = -\arctan\left((\alpha^{'3}_{13} + \cdots)/(\alpha^{''3}_{13} + \cdots)\right) \end{cases} \quad (25)$$

Eq. (23) is used as the temperature analysis model based on harmonic amplitude and phase.

## III. SIMULATION

To validate their accuracy, the analytical expressions for magnetization and magnetic harmonics, which originate from the Fokker-Planck equation with dominant Brownian relaxation, were verified through MATLAB simulations. The relationship between temperature and magnetization response was explored through simulation, and the temperature calculation was performed using the established temperature analysis model (Eq. (23)). The MNPs were subjected to a combined AC and DC field, where the AC field amplitude $\mu_0 H_0$ was 0.002 T, the amplitude of the DC field $\mu_0 H_1$ was 0.001 T, and the AC excitation frequency was 2 kHz. Furthermore, it was assumed that the MNPs had a uniform core size of 18 nm and a hydrodynamic diameter of 64 nm, and the temperatures were set from 310 K to 320 K (310 K, 314 K, and 320 K).

In Fig. 1(a), the black (310 K), red (314 K), and blue (320 K) lines represent the $M$-$H$ curves obtained from the Langevin function. They were compared with the $M$-$H$ curves obtained from the Fokker-Planck equation, where the purple (310 K), yellow (314 K), and brown (320 K) lines represent the $M$-$H$ curves obtained from the Fokker-Planck equation. Besides, the correlation between the magnetization response and temperature was explored. Because of the effect of Brownian relaxation, the $M$-$H$ curves obtained from the Fokker-Planck equation have a hysteresis loop. Figs. 1(b), 1(c), and 1(d) show the $M$-$H$ curves obtained from the analytical expressions ($a_1^1$, $a_1^1 + a_1^3$, $a_1^1 + a_1^3 + a_1^5$) and the Fokker-Planck equation at 310 K, 314 K, and 320 K. The yellow solid lines ($a_1^1$), purple dashed lines ($a_1^1 + a_1^3$) and red dotted lines ($a_1^1 + a_1^3 + a_1^5$) represent the $M$-$H$ curve derived from the analytical expressions. Meanwhile, the insets of Figs 1(b), 1(c), and 1(d) show that the magnetization responses calculated based on analytical expressions with different expansion terms show obvious differences from those from the Fokker-Planck equation. As expansion terms increase, the discrepancy in magnetization responses ($M$-$H$ curves) between the analytical expressions and the Fokker-Planck equation gradually decreases. The analytical expression (dotted lines, $a_1^1+a_1^3+a_1^5$ ) exhibits excellent agreement with the Fokker-Planck results.

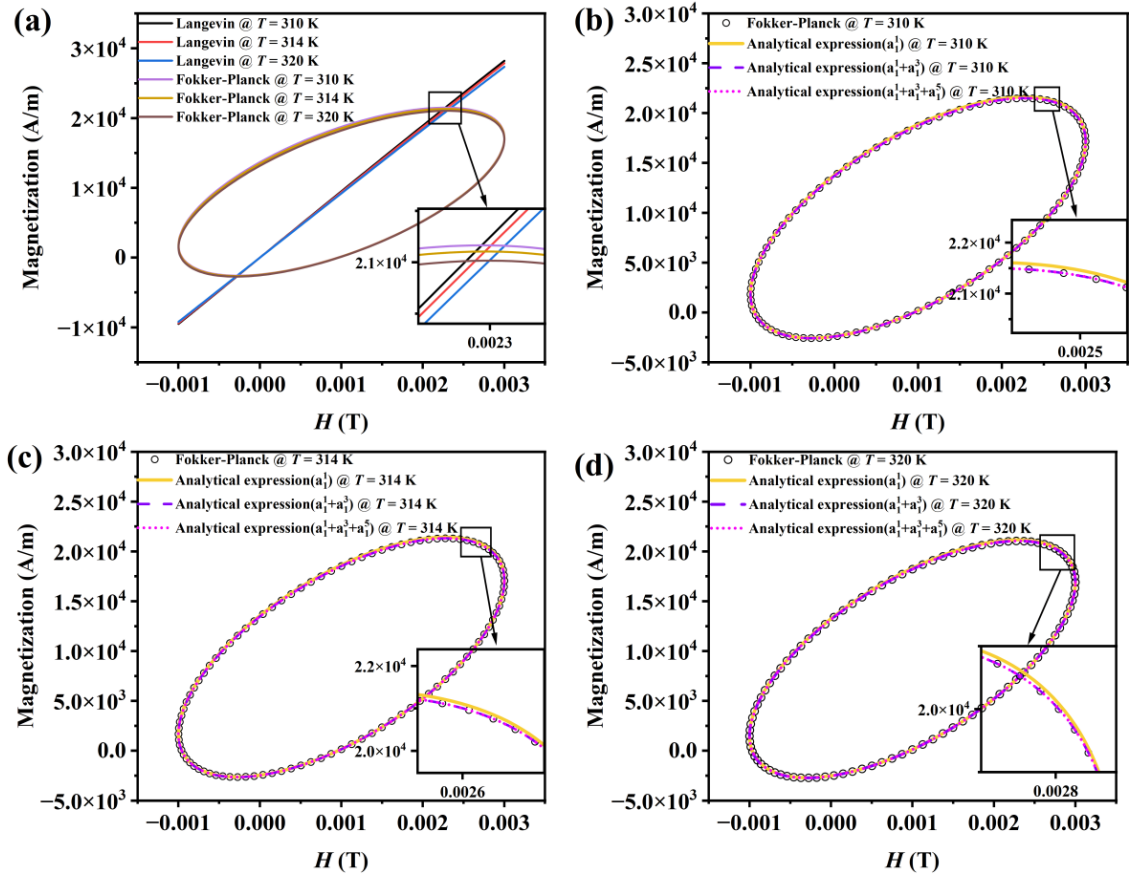

**Fig. 1.** Simulation results for the analytical expressions of the magnetization responses (*M-H* curves) dominated by Brownian relaxation at different temperatures. (a) The trend of magnetization response as a function of temperature. The magnetization responses of analytical expressions with different expansion terms and the Fokker-Planck equation at 310K (b), 314K (c), and 320K (d). The parameters for the simulations are $\mu_0H_0 = 0.002$ T, $\mu_0H_1 = 0.001$ T, $f = 2$ kHz, $d_c = 18$ nm, $d_h = 64$ nm, and $M_s = 200$ kA/m.

Next, simulations were performed at different temperatures, and the MNPs' temperature was calculated using the harmonic amplitude and phase, respectively. It was assumed that the MNPs were subjected to a combined magnetic field consisting of AC field and DC field. The AC field had an amplitude of 0.001 T, with the DC field strength maintained at the same value (0.001 T). The MNPs had a uniform core size of 18 nm and a hydrodynamic diameter of 64 nm. The frequencies of the AC magnetic field were 1270 Hz, 2120 Hz, and 3040 Hz.

As shown in Fig. 2, the magnetic harmonics were calculated from the magnetization responses employing the digital phase-sensitive detection (DPSD) algorithm. The polar coordinate systems were used to represent the harmonic amplitude and phase at frequencies of 1270 Hz (Fig. 2(a)), 2120 Hz (Fig. 2(b)), and 3040 Hz (Fig. 2(c)). The lines in the first and fourth quadrants of the polar coordinate system (clockwise) represent the second, third, and first harmonics at different temperatures, respectively. The length of the lines in the polar coordinate system represents the harmonic amplitudes, and the angles represent the harmonic phase lags. The 0° angle corresponds to the excitation magnetic field direction. As described in Figs.

2(a), 2(b) and 2(c), as the harmonic order increases, the harmonic amplitude decreases, and the phase lag increases clockwise along the excitation magnetic field direction (0°). The effect of Brownian relaxation increases as the excitation frequency increases, which results in more serious phase lag in the clockwise direction. The temperature has an inverse relationship with the effect of Brownian relaxation; accordingly, higher temperatures weaken its influence. Accordingly the harmonic phases gradually align with the excitation magnetic field direction, leading to the phase lag decreasing. The amplitude and phase in Fig.2 were respectively used to calculate the temperature of the MNPs in simulations.

The amplitude-based temperature analysis model was used first to calculate the MNPs' temperature. As shown in Fig. 3, magnetization responses were calculated using Eq. (23), and the DPSD algorithm was applied to obtain the amplitudes of the first, second, and third harmonics at 1270 Hz (Fig. 3(a)), 2120 Hz (Fig. 3(c)), and 3040 Hz (Fig. 3(e)). These amplitudes are represented by black squares (first), red circles (second), and blue triangles (third), respectively. It can be clearly observed that an increase in temperature is accompanied by a monotonic decrease in the harmonic amplitudes. Similarly, increasing the frequency also leads to a reduction in harmonic amplitude. The LM algorithm was employed to calculate the temperature from the harmonic amplitude ratios, and the absolute error of temperature was calculated as the absolute difference between the real temperature and the predicted (estimated) temperature. Figs. 3(b), 3(d), and 3(f) illustrate the temperature estimation errors calculated by the first to second (green squares) harmonics, first to third (purple circles) harmonics, and second to third (yellow triangles) harmonic amplitude ratios at frequencies of 1270 Hz, 2120 Hz, and 3040 Hz. The absolute errors calculated from the second to third harmonic and first to third harmonic amplitude ratios are significantly larger than those from the first to second harmonic amplitude ratio. In the amplitude-based temperature estimation simulation, the temperature error calculated using the first to second harmonic ratio was below 0.0066 K from 310 K to 320 K.

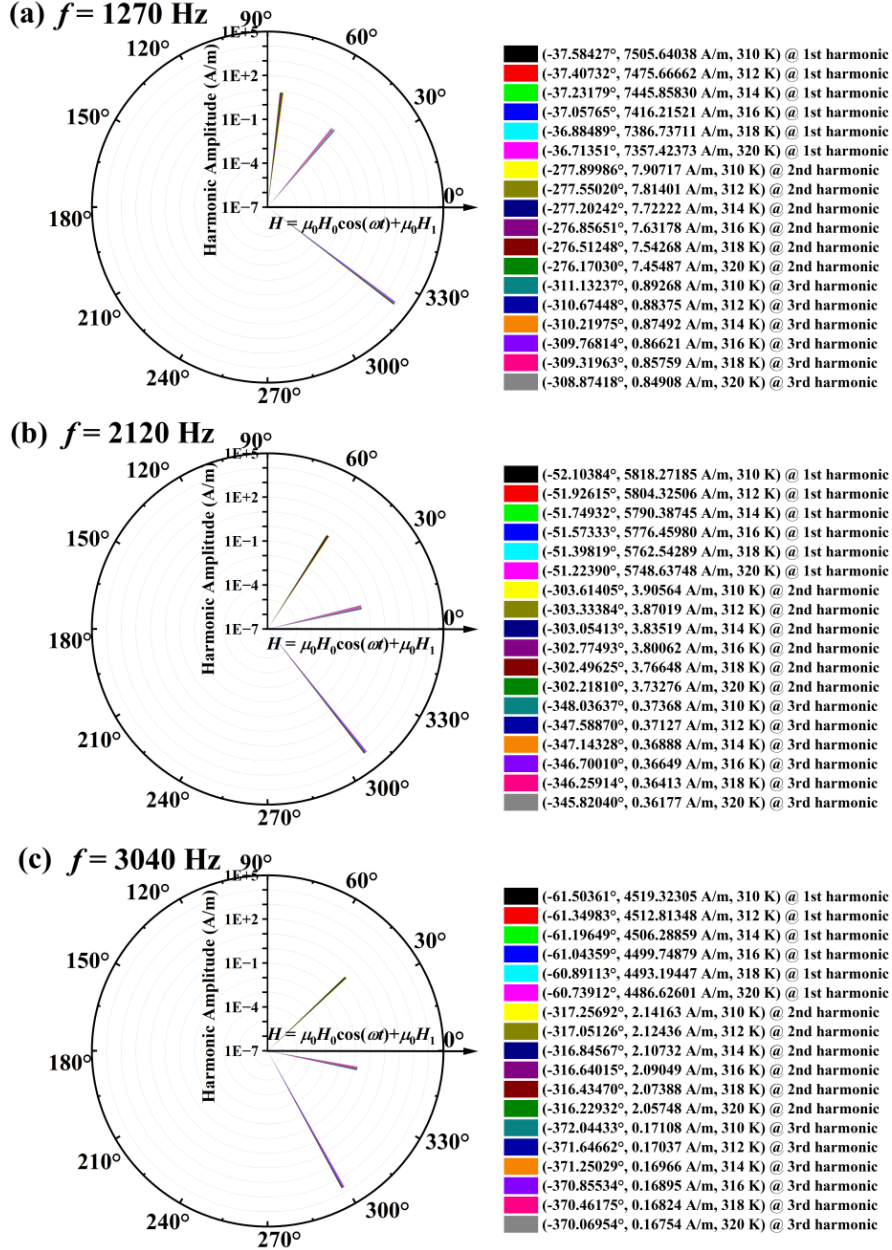

**Fig. 2.** Polar coordinate systems of magnetization harmonics at different frequencies. The amplitude and phase of the first, second and third magnetic harmonics obtained from the analytical expressions at 1270 Hz (a), 2120 Hz (b) and 3040 Hz (c). The parameters for the simulations are $\mu_0 H_0 = 0.001$ T, $\mu_0 H_1 = 0.001$ T, $d_c = 18$ nm, $d_h = 64$ nm, and $M_s = 200$ kA/m.

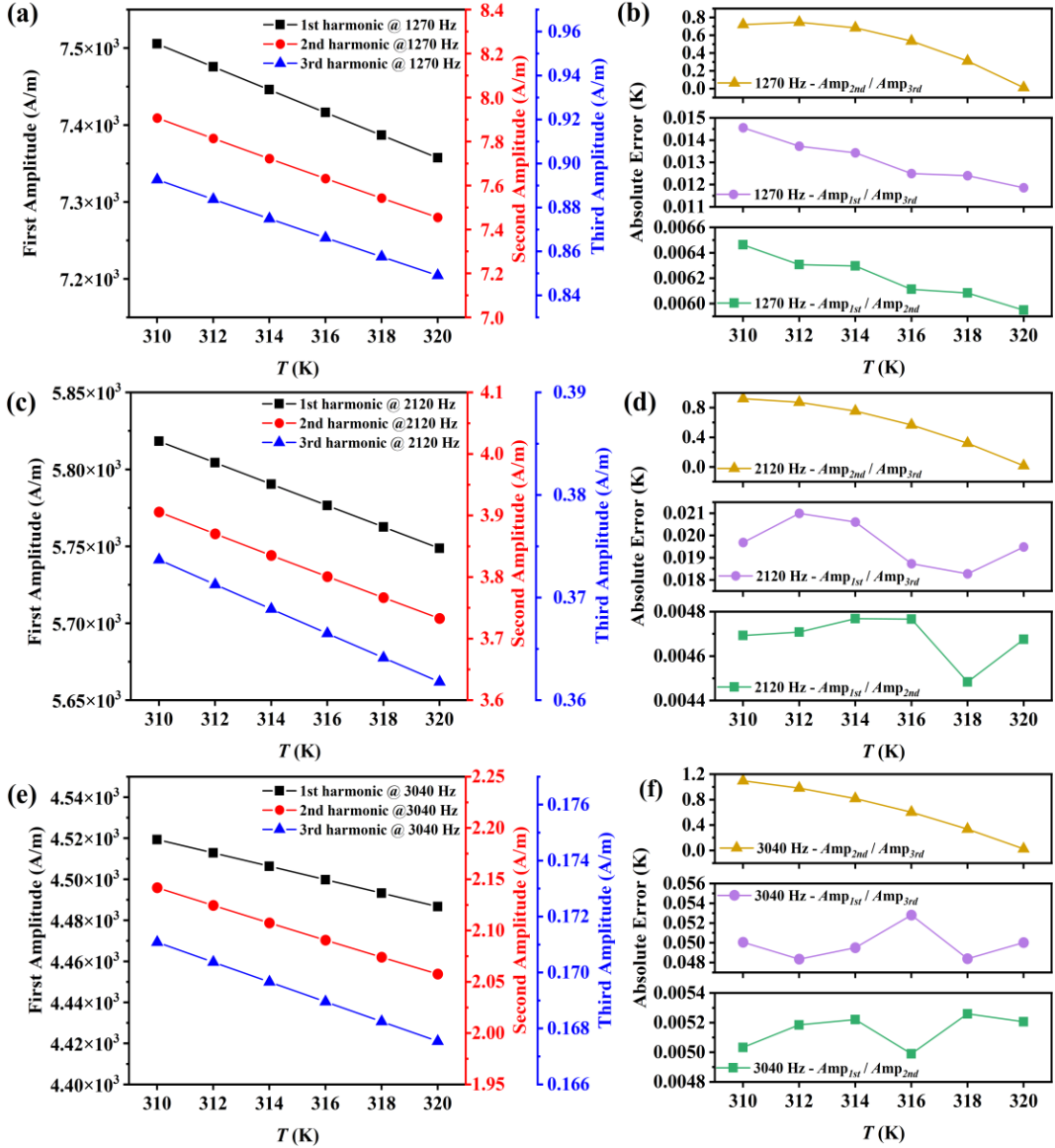

**Fig. 3.** Simulation results of temperature estimation based on harmonic amplitude. The first (black squares), second (red circles), and third (blue triangles) harmonic amplitudes were obtained at 1270 Hz (a), 2120 Hz (c), and 3040 Hz (e). The temperature estimation errors based on the amplitude ratios of the first to second (green squares) harmonics, first to third (purple circles) harmonics, and second to third (yellow triangles) harmonics were obtained using Eq. (23) at 1270 Hz (b), 2120 Hz (d), and 3040 Hz (f), respectively.

Then, the phase-based temperature analysis model was applied to calculate the MNPs' temperature. As shown in Figs. 4(a), 4(c), and 4(e), the phases of the first, second, and third harmonics were computed at 1270 Hz, 2120 Hz, and 3040 Hz, and were represented by black squares, red circles, and blue triangles, respectively. As the harmonic order increases, the phase lag increases. It can be clearly observed that the lag of the harmonic phase obtained from the analytical expression decreases with increasing temperature. This means that as the temperature increases, the Brownian relaxation time ($\tau_{B,0}$) decreases. This leads to the harmonic phase shifting toward the excitation magnetic field direction, thereby reducing the phase lag. The harmonic phase lag obtained from the magnetization response gradually increases with the excitation frequency. This means the harmonic phase exhibits a clockwise shift relative to the excitation magnetic field direction. Figs. 4(b), 4(d), and 4(f) depict the temperature estimation errors based on the first, second, and third harmonic phases at 1270 Hz, 2120 Hz, and 3040 Hz, which were represented by green squares, purple circles, and yellow triangles, respectively. The first, second, and third harmonic phases were used to calculate temperature. The temperature estimation error calculated from the first harmonic phase was much smaller than those from the second and third harmonic phases. In this simulation, the absolute error of the temperature estimated using the first harmonic phase is below 0.000041 K from 310 K to 320 K.

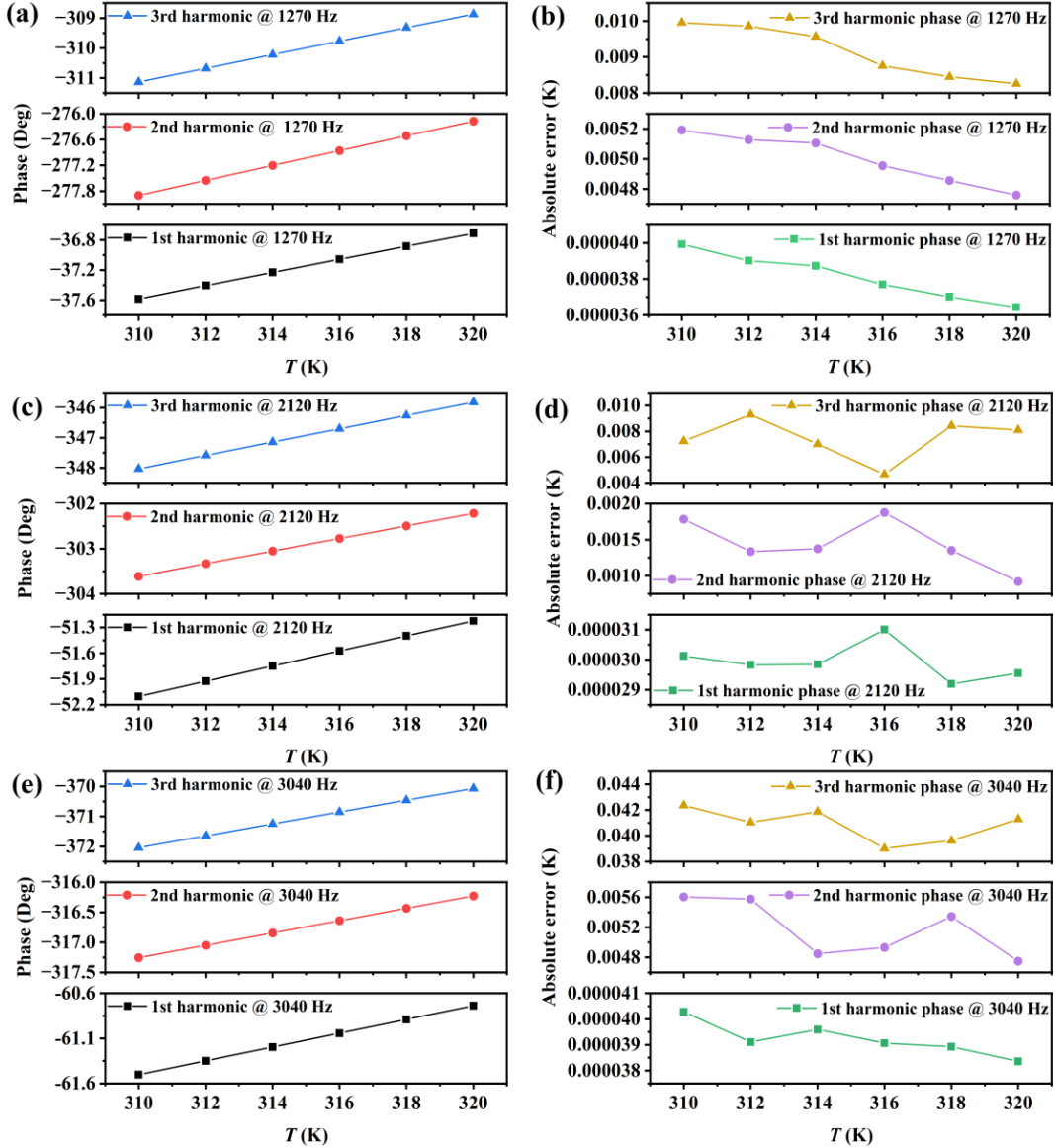

**Fig. 4.** Simulation results of temperature estimation based on harmonic phase. The phases of the first (black squares), second (red circles), and third (blue triangles) harmonics were obtained at 1270 Hz (a), 2120 Hz (c), and 3040 Hz (e). The temperature estimation errors based on the first harmonic phase (green squares), second harmonic phase (purple circles), and third harmonic phase (yellow triangles) were obtained using Eq. (23) at 1270 Hz (b), 2120 Hz (d), and 3040 Hz (f), respectively.

## IV. EXPERIMENT AND RESULTS

The experiments used commercially available MNPs (SHP15 and SOR30) from Ocean NanoTech (USA), and MNPs (EMG1300) from Ferrotec (Japan). The SHP15 MNPs, with a specified iron concentration of 5 mg/mL, are supplied in deionized water. The SOR30 MNPs, with a specified iron concentration of 25 mg/mL, are supplied in chloroform. The EMG1300 MNPs are used in the form of a solid powder. In this experiment, the frequencies of the AC magnetic field were 1270 Hz, 2120 Hz, and 3040 Hz, with an amplitude of 0.001 T, while the DC magnetic field was the same value (0.001 T).

The first and second harmonic amplitudes were used to calculate the temperature of the MNP-based samples (SOR30, SHP15, and EMG1300). As shown in Figs. 5(a), 5(c), and 5(e), the first (red) and second (black) harmonic amplitudes of three MNPs were measured at frequencies of 1270 Hz (squares), 2120 Hz (circles), and 3040 Hz (triangles). An increase in frequency results in a reduction of the harmonic amplitude in MNPs. Similarly, with an increase in temperature, the harmonic amplitudes also monotonically decrease. Figs. 5(b), 5(d), and 5(f) show the temperature errors calculated using the LM algorithm. The purple squares (1270 Hz), yellow circles (2120 Hz), and green triangles (3040 Hz) represent the MNPs' temperature errors. The maximum temperature errors for the three MNP samples were below 0.0116 K at 1270 Hz, below 0.0142 K at 2120 Hz, and below 0.0155 K at 3040 Hz. The reason may be that an increase in frequency result in a reduction in the first and second harmonic amplitudes in MNPs. The noise level remains consistent at different frequencies, which results in a lower signal-to-noise ratio of the harmonics and larger calculation errors.

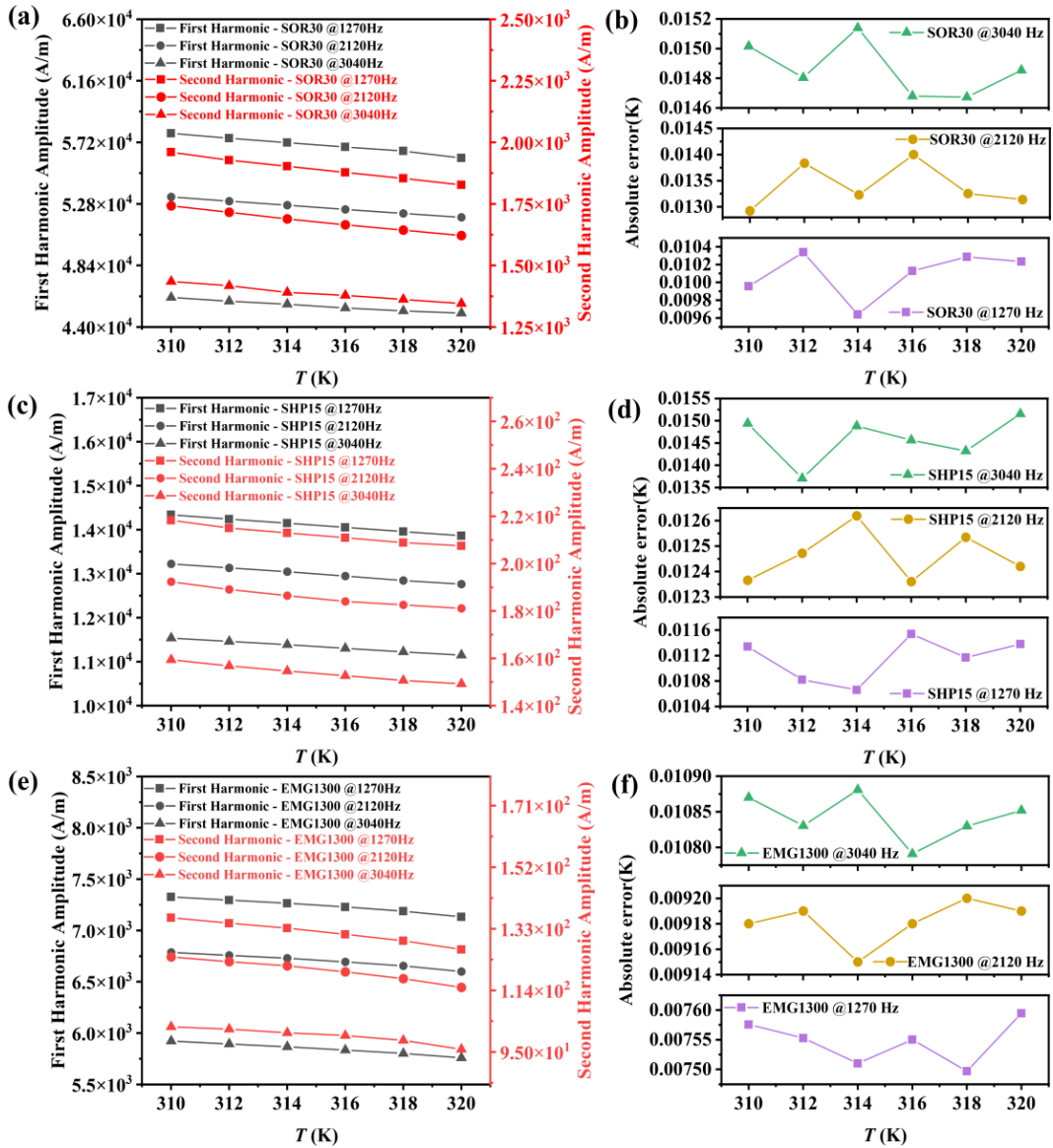

**Fig. 5.** Experimental results of temperature estimation based on harmonic amplitudes. The amplitudes of the first and second harmonics for SOR30 (a), SHP15 (c), and EMG1300 (e) were obtained at frequencies of 1270 Hz, 2120 Hz, and 3040 Hz, respectively. Temperature errors from 310 K to 320 K were obtained for SOR30 (b), SHP15 (d), and EMG1300 (f) at frequencies of 1270 Hz, 2120 Hz, and 3040 Hz, respectively.

The first harmonic phase was then used to calculate the temperature of three MNP-based samples (SOR30, SHP15, and EMG1300). As shown in Figs. 6(a), 6(c), and 6(e), the first harmonic phases of MNPs were measured at frequencies of 1270 Hz (black squares), 2120 Hz (red circles), and 3040 Hz (blue triangles). An increase in temperature causes a reduction in the Brownian relaxation time, which in turn leads to a decrease in the lag of first harmonic phase. Besides, as the excitation frequency increases, the Brownian relaxation proportion also increases. This means that the lag of the first harmonic phase becomes serious with the increase of excitation frequency. Figs. 6(b), 6(d), and 6(f) describe the absolute error of the temperature estimation at different frequencies for the MNP samples. The purple squares, yellow circles, and green triangles represent the absolute error of the temperature calculations at frequencies of 1270 Hz, 2120 Hz, and 3040 Hz. The maximum temperature error for the MNPs is approximately 0.0171 K at a frequency of 1270 Hz. As the frequency increases to 2120 Hz, the absolute error increases to around 0.0195 K, and at 3040 Hz, the maximum temperature error further increases to around 0.0218 K.

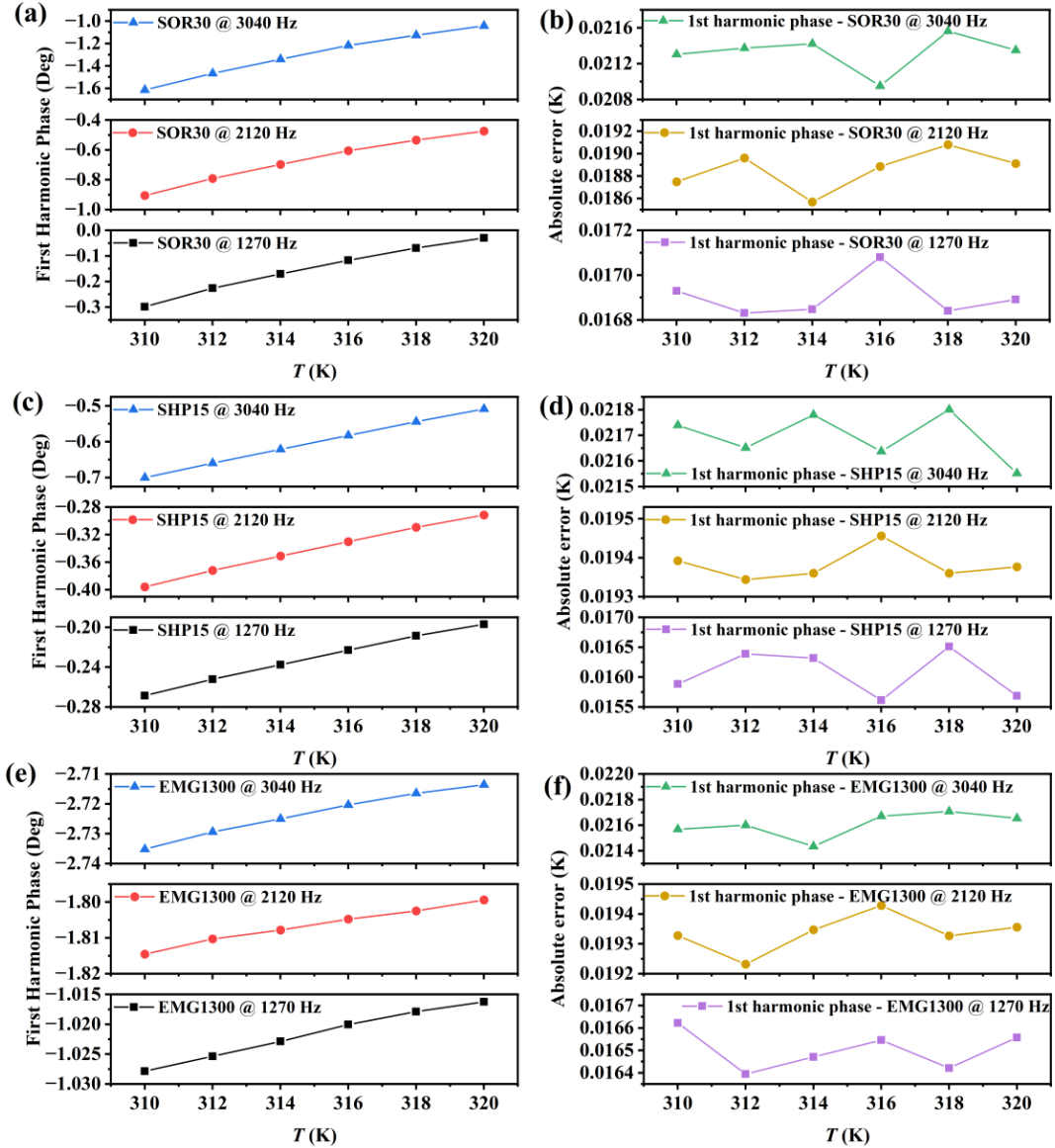

**Fig. 6.** Experimental results of temperature estimation based on the first harmonic phases. The phases of the first harmonics for SOR30 (a), SHP15 (c), and EMG1300 (e) were obtained at 1270 Hz, 2120 Hz, and 3040 Hz, respectively. Temperature errors from 310 K to 320 K were obtained for SOR30 (b), SHP15 (d), and EMG1300 (f) at 1270 Hz, 2120 Hz, and 3040 Hz, respectively.

## V. DISCUSSION

From the simulation, it can be observed that the truncation error between the analytical expression derived from the perturbation method and the Fokker-Planck equation is dependent on the number of expansion terms. The truncation error generally decreases as the expansion terms increase, because the addition of more expansion terms can improve the approximate accuracy of the analytical expression for the Fokker-Planck equation. For Figs 3(b) and 4(b), the error of the estimated temperature at a frequency of 1270 Hz shows a monotonically decreasing trend. The reason may be that at an excitation frequency of 1270 Hz, although Brownian relaxation still dominates, its proportion is relatively low, requiring the addition of more expansion terms in the analytical expression. However, the more expansion terms there are, the more complex the temperature analysis model is, and the longer the calculation time is. As a result, there is a compromise between computational cost and accuracy requirements.

In the phase-based temperature analysis model, the harmonic phase is unrelated to the MNP volume fraction, less sensitive to temperature changes, and highly susceptible to noise interference. On the other hand, the amplitude-based temperature analysis model requires measuring the amplitudes of two harmonics and using the harmonic amplitude ratio to eliminate the MNPs' volume fraction. The amplitude is more sensitive to temperature changes, and the noise has less effect on the harmonic amplitude. Therefore, although the amplitude-based temperature analysis model is time-consuming, it requires less precision in the experimental equipment. In contrast, the phase-based temperature estimation method allows for faster temperature estimation but requires the experimental equipment to have higher measurement precision.

The current analytical expressions are applicable only to MNPs dominated by Brownian relaxation. These analytical expressions do not consider more complex scenarios where both Brownian and Néel relaxations coexist, or where Néel relaxation becomes dominant at high frequencies of AC magnetic fields. To improve the applicability of the analytical model, we may focus on extending the current expressions to incorporate the effects of Néel relaxation, especially under high-frequency conditions where its influence is significant. Developing a hybrid model that describes both Brownian and Néel relaxation mechanisms would enable a more comprehensive characterization of MNPs' magnetization dynamics across a broader range of particle sizes, excitation magnetic field strength, and frequencies. Such a hybrid model could significantly improve the accuracy of temperature estimates, particularly in complex magnetic field conditions, enhancing temperature estimation in MNPH applications.

## VI. CONCLUSION

In this paper, the perturbation method was used to derive the analytical expression for the Fokker-Planck equation dominated by Brownian relaxation under a combined AC and DC magnetic field. By comparing the magnetization responses derived from the Fokker-Planck equation and analytical expressions, we demonstrate that the analytical expressions can precisely describe the magnetization dynamics of the Fokker-Planck equation. A temperature estimation model based on harmonics is then established. Temperature calculation was performed using the LM algorithm based on amplitude ratio of first to second harmonic and phase of the first harmonic, respectively. The experimental portion of this study demonstrated that the absolute error of temperature based on the amplitude ratio of the first to second harmonic is below 0.0151 K, and the error based on the first harmonic phase is below 0.0218 K from 310 K to 320 K. These results highlight the potential of the proposed model to significantly enhance temperature measurement accuracy and its applicability in more complex MNPH and MNP imaging scenarios.

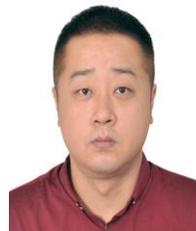**Zhongzhou Du** received the B.S. degree from henan university of science and technology, China, 2007, received the M.S. and Ph.D degrees from the School of Artificial Intelligence and Automation, Huazhong University of Science and Technology (HUST), Wuhan, China, in 2012, 2015, respectively. In 2015, he joined Zhengzhou University of Light Industry, where he is a Lecturer. In 2018, he received an award from the Japan Society for the Promotion of Science (JSPS), and was engaged in research work as a postdoctoral fellow in Kyushu University, 2018-2020. In 2020 he joined Zhengzhou University of Light Industry again, where he is an associate professor with the School of Computer and Communication

Engineering. His current research interests include noninvasive temperature measurement using MNPs and magnetohydrodynamic simulation and computation.

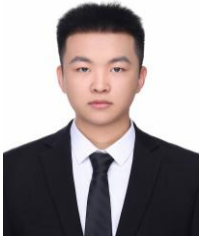

**Wenze Zhang** was born in Henan, China, in 2000. He is currently pursuing the master's degree with Zhengzhou University of Light Industry. His research interest includes magnetic temperature measurement technology

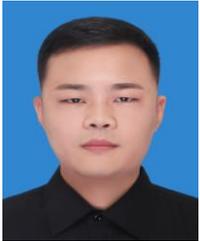

**Yi Sun** joined the Kyushu University (Japan) at Fukuoka for a Ph.D. degree. He received the M.S. and B.S. degree in mechanical engineering from Zhengzhou University of Light Industry. His research interests include magnetic nanoparticle characterization, magnetometer and magnetic nanothermometer, as well as their applications in biomedical, biological and industrial applications.

**Na Ye** received the B.S. and M. S. degree from henan university of science and technology, China, in 2007, 2011, respectively. In 2015, she joined Zhengzhou University of Light Industry, where he is a Lecturer with the School of Computer and Communication Engineering. Her current research interests include noninvasive temperature measurement using MNPs and digital signal processing (DSP).

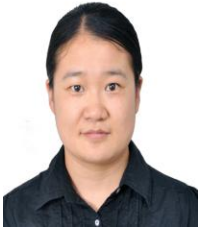

**Yong Gan** received the Ph.D degrees from the computer science and technology, Xi'an Jiaotong University, Xi'an, China, in 2006. His research interests include magnetic temperature measurement and magnetohydrodynamic computation.

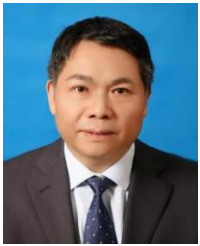

**Pengchao Wang** was born in Hunan, China. He is currently pursuing the master's degree with Zhengzhou University of Light Industry. His research interests include magnetic temperature measurement technology and simulation.

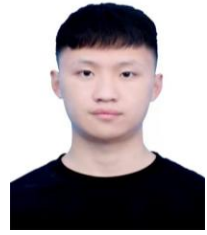

**Xinwei Zhang** was born in Henan, China, in 2004. He is currently pursuing the bachelor's degree with Zhengzhou University of Light Industry. His research interests include magnetohydrodynamic simulation and computation.

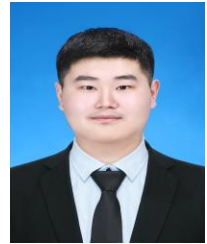

**Yuanhao Zhng** was born in Henan, China in 2003. He is currently pursuing the master's degree with Zhengzhou University of Light Industry. His research interests include magnetohydrodynamic simulation.

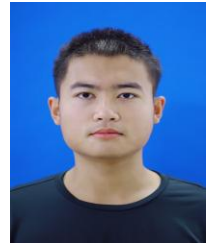

**Shijie Han** received the B.S. degree in measurement and control technology and instrumentation from Huazhong University of Science and Technology, Wuhan, China, in 2022. He is currently pursuing the Ph.D. degree in control science and engineering with Huazhong University of Science and Technology.

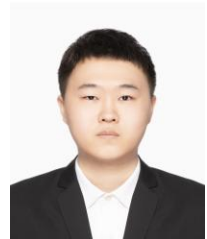

**Haochen Zhang** received the bachelor's degree in the major of Electrical Engineering and Automation of Xi'an University of Science and Technology, China, Xi'an, in 2015. In 2019, he joined the Kyushu University (Japan) at Fukuoka for a master degree. His research interests include bioapplication using magnetism, magnetic immunoassys. Currently he is an Active graduate student in Kyushu University.

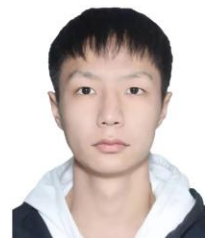

**Haozhe Wang** is currently pursuing the Ph.D. degree in department of electrical and electronic engineering with Kyushu University. His current research interests include detection technology and noninvasive temperature measurement using MNPs.

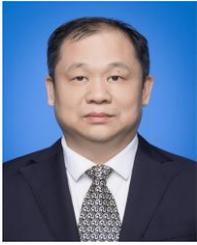

**Wenzhong Liu** (Member, IEEE) received his B.S., M.S., and Ph.D. degrees from School of Artificial Intelligence and Automation, Huazhong University of Science and Technology, Wuhan, China, in 1997, 2000, and 2004, respectively. In 2000 he joined the Huazhong University of Science and Technology where he is a Lecturer. Currently he is a full professor in School of Artificial Intelligence and Automation, Huazhong University of Science and Technology. He specializes in weak signal detection, focusing on issues related to characterization of magnetic nanoparticles (MNP), temperature estimation using MNP and MNP imaging. His current research interests include non-invasive temperature imaging using MNP, and characterization of MNP.

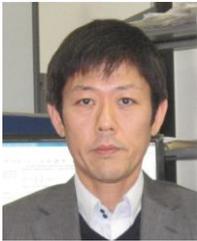

**Takashi Yoshida** (Member, IEEE) is a professor from institute of superconductors science and systems in Kyushu University (Japan) at Fukuoka. His research interests include magnetic nanoparticle characterization, magnetometer and magnetic particle imaging, as well as their applications in biomedical, biological and industrial applications.